\documentclass[twocolumn,showpacs,preprintnumbers,amsmath,amssymb]{revtex4}
\usepackage{graphicx}
\usepackage{dcolumn}
\usepackage{bm}

\begin{document}

\title{Molecular dynamics simulation of polymer insertion into lipid bilayers}
\author{Mitsuharu Okazaki}
\author{Tomoki Watanabe}
\author{Naohito Urakami} 
\author{Takashi Yamamoto}
\email{yamamoto@mms.sci.yamaguchi-u.ac.jp}
\affiliation{ Faculty of Science, Yamaguchi University, Yamaguchi 753-8512, Yamaguchi, Japan}
\date{\today}

\begin{abstract}
 Relatively short peptides, such as toxins and antimicrobial-peptides, are known to insert themselves into cell membranes.  On the basis of simple bead-spring models for the membrane lipids, the peptide, and water, detailed processes of the peptide insertion is investigated by molecular dynamics simulation; our special concern is in the highly cooperative motions of membrane lipids and the peptide.  Our model lipid has a head group of three hydrophilic beads and a tail of seven hydrophobic beads, while the model peptide is a block-copolymer made of hydrophilic and hydrophobic blocks, with total length of 200 beads.  In addition, each water molecule is represented by a single bead which has considerably larger interaction energy.  We first confirm that our present lipid model can support spontaneous formation of bilayers in water. Then we place the model peptide near the bilayers and monitor the microscopic process of the adsorption, insertion, and translocation of the peptide.  When the peptide molecule is set free to interact with the membrane, the hydrophobic blocks of the peptide strongly favor intimate contact with the lipid tails.  This strong attractive interaction gives rise to a severe membrane perturbation, which leads to the formation of a pore, though short-lived, in the membrane.  It is found that the side-surface of the pore is almost covered with the hydrophilic heads of the lipids, which seems to help the hydrophilic blocks of the peptide translocate across the membrane.  We also monitor the rate of flip-flop inversion of the directions of lipids, and we find it markedly increases during the peptide contact and insertion.  The peptide insertion/translocation, the formation of membrane pore, and the flip-flop motions of the lipids are thus found to be closely interconnected. 
\end{abstract}

\maketitle

\section{ \label{sec:level1} Introduction }
Lipid bilayers are common frameworks of various biological membranes.  Microscopic structures of the bilayers, molecular level understanding of their formation, various morphologies they exhibit, and their interactions or fusions, are all of fundamental importance in the studies of cell membranes \cite{membrane}.  Most of the exquisite functions of the cell membranes are, however, carried out by cooperative work of lipids and membrane proteins: receptors, ion channels, etc.  In fact, proteins are actively moving even within the cell by inserting into or translocating across membrane systems between various organelles \cite{cell}.

Peptides and proteins usually accomplish insertion or translocation either assisted by complex proteinacious machinery or through fusion of membranes \cite{cell}.  Relatively small peptides such as toxins and antimicrobial peptides, however, are known to insert themselves by physicochemical processes, in which four-stage or two-stage models have been proposed for the insertion and reorganization \cite{engel-1,engel-2,engel-3}.  The peptides first get adsorbed to the membrane surface where they change their conformation into $\alpha$-helix due to large hydrophobic environment at the surface. Then they start insertion, whose molecular mechanism is least understood, followed by aggregation into an organized structure.  The driving force easily think of is the hydrophobic nature of the peptide segments.  In fact peptide insertions are usually discussed by use of the hydrophobicity scale of the component amino acids \cite{white,blanco}.  Other factors that control insertion such as membrane curvatures are also very intriguing \cite{baum-1,baum-2}.  We should have it in mind that the insertion process depends sensibly on the primary structure of the peptide such as the lengths of the hydrophilic and hydrophobic sequences as well as on the properties of the matrix lipid membrane such as surface charge distributions and their imbalance between extracellular and cytoplasmic sides.

  Formations of various complexes in systems of water, surfactants, and amphiphilic block-copolymers are also of great interest from other points of view.  The block-copolymers are known to stabilize bilayer structures of the L$_{\alpha}$ phase of lipids \cite{lig}; it is also found that small addition of the block-copolymer drastically increase the emulsification capacity of the surfactants \cite{endo}.  Of great biological and pharmaceutical interest is the finding that hydrophilic polymers with hydrophobic anchor greatly improve the stability the conventional liposome opening up new possibilities in drug delivery systems \cite{lasic}.

  Despite large academic interest and potential applications, complex cooperative actions of peptides (block-copolymers) and lipids that take place in local nanometer-sized space are often beyond the reach of experiments.  Molecular simulation is now emerging as a very promising tool to study such complicated molecular processes.  Molecular simulations of membrane proteins especially of their equilibrium structure and dynamics supporting biological functions have been investigated intensively \cite{md}, but there are only few reports on the molecular simulation studies of peptide insertion leading finally to the folded structures \cite{milik,baum-3,maddox,bern}.  Successful works were made mostly by modeling the membrane as a continuum or an aggregation of rigid cylindrical rods, in which detailed cooperative motions of the lipids and the peptide were not questioned \cite{milik,baum-3,maddox}.

   We here consider the lipid molecule with a water-like hydrophilic head group and a polymethylene-like hydrophilic tail, while the peptide is modeled as a block-copolymer made of hydrophobic and hydrophilic blocks of discrete hydrophobicity scales, though real peptides are made of amino acids having wide spectrum of hydrophobicity.  Once such an approach is found successful, however, the elaboration of the model into more realistic ones may be straightforward. 
  
\section{ Model and method }
We will investigate the molecular mechanism of insertion of a model peptide into lipid bilayers by use of molecular dynamics (MD) simulation. We here adopt psudoatomic models for water, lipids, and peptide; characteristic length scale is of atomic dimension but the molecular structures are considerably simplified to accelerate computation. Lipid molecules that constitute the cell membrane are amphiphilic molecules that have both hydrophilic head groups and the hydrophobic tails.  We consider a model lipid with a head group made of three hydrophilic beads (P-beads), and a tail made of seven hydrophobic beads (H-beads) (Fig.1a). 
 On the other hand the peptide is modeled as a block-copolymer made of hydrophilic blocks of the P-beads and hydrophobic blocks of the H-beads (Fig.1b).  The block-copolymer is assumed to be made of 200 beads and have a sequence P$_5$(H$_{15}$P$_{15}$)$_6$H$_{15}$; the length of the hydrophobic blocks of 15 H-beads is so selected to give nearly the same length as the thickness of the bilayer in order to facilitate incorporation into the membrane.  The H-bead is here considered to be the usual united atom CH$_2$ of mass 14, while the P-bead has also the same mass and interactions as the H-bead with only difference in the affinity to water.  Each water molecule is represented by a single bead (W-bead) which has much larger interaction energy as described below.  

\begin{figure}
\includegraphics[width=.9\linewidth]{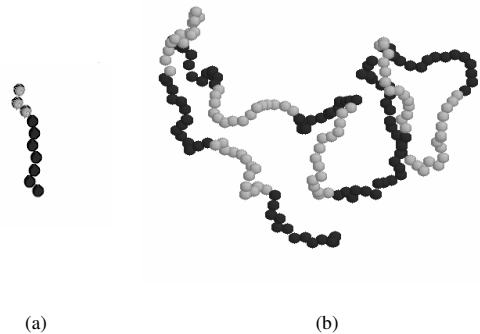}
\caption{ (a) The model lipid is made of a head-group (three hydrophilic atoms P$_3$: grey beads) and a tail (seven hydrophobic atoms H$_7$: black beads). (b) The model peptide is a copolymer made of hydrophilic (grey) and hydrophobic (black) blocks: P$_5$(H$_{15}$P$_{15}$)$_6$H$_{15}$ }
\end{figure}

 The lipid and peptide molecules are assumed to be fully flexible, where following bond-stretching potential is only imposed neglecting the energies of distortions in the bond angles and dihedral angles,

\begin{eqnarray}
 U_{bond} = \frac{1}{2} k_b (r_{i,i+1} - r_0 )^2
\end{eqnarray}

where $r_{i,i+1}$ is a length of the i-th bond, $r_0$ is the equilibrium bond length, and  $k_b$ is the spring constant.  Following non-bonded interactions are considered between two beads of the same chain more than two bonds apart, and between those of different chains.  Between the beads of the similar hydrophobicity (P-P, H-H, W-W, and P-W), we assume the conventional Lennard-Jones interactions of the type,

\begin{eqnarray}
 U_{vdW} = 4 \epsilon [(\frac{\sigma}{r})^{12} - (\frac{\sigma}{r})^6]
\end{eqnarray}

, while between the hydrophilic and hydrophobic beads (H-P, and H-W) we impose soft core repulsion

\begin{eqnarray}
 U_{sc} =  4 \epsilon (\frac{\sigma}{r})^9
\end{eqnarray}

Van der Waals attraction is here cut-off at a rather long distance   as to assure the peptide's approach to the membrane. The water molecules are modeled as beads but with four times larger attractive interaction $\epsilon$ in equation (2) considering approximately four time higher boiling point of water than that expected for hypothetical CH$_2$ liquid; the hydrophilic head of the lipid made of three beads is so constructed to have nearly the same total interaction with the water molecule.  Present molecular model of lipid is largely based on a paper by Goetz and Lipowski \cite{goets-1,goets-2}, but the molecular model is slightly fine grained to assure better correspondence to the usual hydrocarbon chain which we have hitherto been studying.  The potential parameters used in the present simulation is given in tabled I.  
 
\begin{table}
\caption{ Molecular parameters used in the simulation }
\begin{ruledtabular}
\begin{tabular}{lcr} 
 parameter & in absolute units & in reduced units \\
\hline
 $l_0$           &   0.152 (nm)                   &   0.4     \\
 $k_b$           &  $3.46 \times 10^7$ (J/mol/nm$^2$) &   9993    \\
 $\sigma_{lipid}=\sigma_{peptide}$  &   0.38 (nm)      &   1       \\
 $\sigma_{water}$     &   0.33 (nm)                    &   0.88    \\
 $\epsilon_{lipid}=\epsilon_{peptide}$ &  500 (J/mol)  &   1       \\
 $\epsilon_{water}$   &   2000 (J/mol)                 &   4.0     \\
 $m$            &   $14.0 \times 10^{-3}$ (kg/mol) &    1      \\
\end{tabular}
\end{ruledtabular}
\end{table}

  We made MD simulations under NTV ensemble with MD cell size 20$\sigma$ in each direction and periodic boundary condition.  We placed within the MD cell 305 lipids each made of 10 beads and a peptide molecule of 200 beads together with surrounding 3675 water molecules.   We solved Newton's equations of motion by the leapfrog method with temperature control at 300K by velocity scaling.  We first confirmed that our present lipid model supports spontaneous formation of a bilayer membrane in water.  Then we placed a model peptide molecule near the membrane and monitored the process of adsorption and insertion into the membrane.

\section{Results and Discussions}

\subsection{Spontaneous membrane formation}
 The mesoscopic structure formation of lipids is a result of subtle balance of interactions among the hydrophilic heads, the hydrophobic tails, and water molecules.  Depending on the density of lipids, PH of the solvent, etc. various morphologies appear: micelles, vesicles, membranes, etc \cite{membrane}.  We first examined whether our molecular model supports the spontaneous formation of bilayer membrane in water.  It is readily noticed that, due to the applied periodic boundary condition, the membrane formation is possible only when the natural lateral width of the membrane fits the size of the MD cell, otherwise they will form spherical micelles, cylindrical micelles, etc \cite{goets-1}.  By proper choice of the MD cell size (20$\sigma$) and the number of lipids (305 chains), we could confirm a gradual formation of MD-cell-spanning bilayers out of random mixture of lipids and waters.  In all the simulations that follow, we adopted this favorable lipid number and cell size that allow the membrane formation.

  Typical snapshot of the bilayers (Fig.2a) obtained after 1.0 ns of simulation at 300K shows that the observed membrane has large dynamical disorder that makes the quick response to external perturbation possible.  The density profiles of the atoms are given in Fig.2b, where both hydrophilic and hydrophobic atoms show rather wide distributions but still with a definite hydrophobic core. 
\begin{figure}
\includegraphics[width=.9\linewidth]{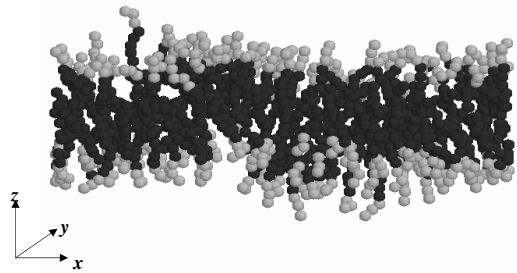}
\includegraphics[width=.9\linewidth]{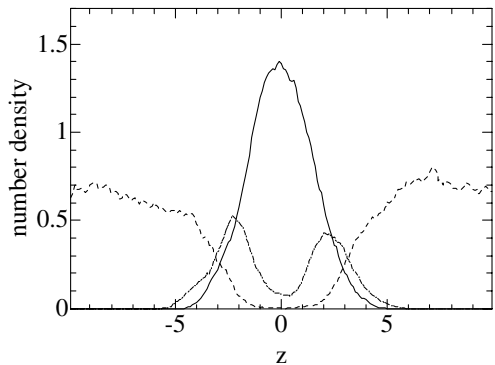}
\caption{(a) A snapshot of the lipid bilayers in water spontaneously formed from a random mixture of water and lipid molecules at 300K. The z axis is taken perpendicular to the membrane.  (b) The reduced number density (average number of atoms per unit volume ), projected along the x-y plane, of the hydrophilic atoms (dash-dot) and hydrophobic atoms (solid) together with water (dash); the abscissa is also reduced by unit.}
\end{figure}

\subsection{Peptide conformation before partitioning }
 A free amphiphilic block-copolymer in water is considered to have specific isotropic structure.  Before discussing insertion into the membrane, we give a glimpse at the conformation of our model peptide in water.  Shown in Fig.3 is a typical equilibrium conformation of the model peptide made of sequential hydrophilic and hydrophobic blocks.  The peptide is stretching six hydrophilic block loops out into the water phase, while the hydrophobic blocks are segregated and shrunk to form spherical core.  The amphiphilic peptide molecule is thus imagined as a unimolecular micelle with hydrophobic core segments surrounded by hydrophilic loops. 
\begin{figure}
\includegraphics[width=.9\linewidth]{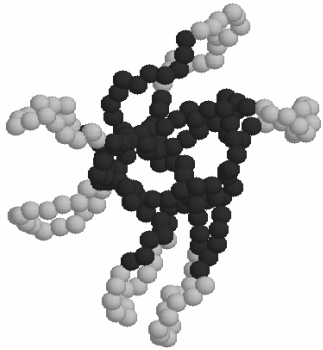}
\caption{ Typical chain conformation of the model peptide free from interactions with the membrane.  The molecule is deemed as a unimolecular micelle stretching seven hydrophilic arms out into the water phase while the hydrophobic blocks are cramped to form a central core region. }
\end{figure}
 This characteristic initial conformation of our model peptide in water has been profoundly distorted when it comes close to touch the bilayer membrane.  The isotropic distribution of the hydrophilic loops around the hydrophobic core is disrupted, and the hydrophobic core gets bared of the hydrophilic overcoat resulting in strong interactions with the hydrophobic tails of the membrane.  This also gives larger perturbation to the integrity and the planarity of the membrane, which finally leads to a successful translocation of the hydrophilic blocks of the peptide against the hydrophobic barrier in the membrane core as described below (Fig.4).

\subsection{Peptide insertion into the membrane}
 When the peptide molecule is set free to interact with the membrane, the interaction gives strong perturbation to the structures of both peptide and membrane near the point of contact. The hydrophobic blocks of the peptide strongly favor intimate contact with the lipid tails when they come closer than the interaction cut-off 4$\sigma$, while the attraction between the hydrophilic blocks of the peptide and the lipid heads is moderated by the intervening water molecules.

 The core of the unimolecular micelle peptide attracts lipid tails and tends to form even larger micelle resulting in depletion of the lipid molecules around the contact point (Fig.4c).  Such large perturbation to the membrane often gives rise to the creation of a pore, though short-lived, in the membrane.  Configuration of the chains around the pore is seen in Fig.4d, where lipid molecules tend to lie horizontally with their hydrophilic heads pointing outward into the water phase.  We can notice that the side-surface of the pore is almost covered with the hydrophilic heads of the lipids, and this is considered to help the hydrophilic blocks of the peptide translocate across the membrane.  The opening of the pore is only temporary and the membrane integrity is readily recovered in a very short time (Fig.4e).  Finally the peptide acquires its stable position forming neat folded conformation within the membrane (Fig.4f). 
\begin{figure}
\includegraphics[width=.9\linewidth]{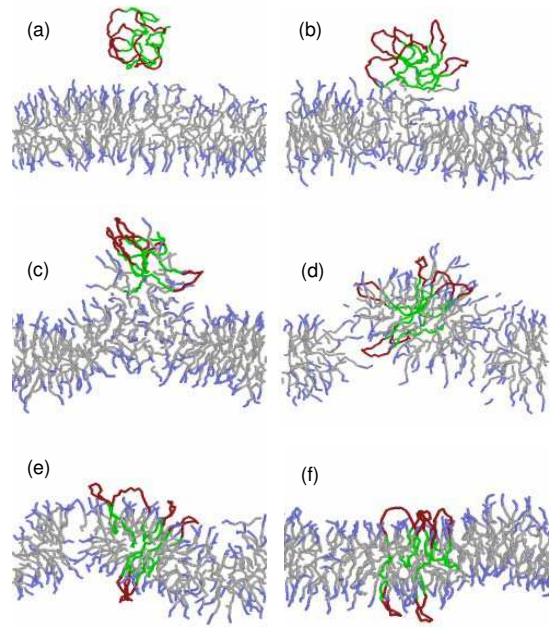}
\caption{Typical trajectories at 300K of the peptide successfully translocated the bilayers: (a) the initial state at 0.1 ns, (b) start contacting at 0.2 ns, (c) uptake of lipid tails at 0.3 ns, (d) pore formation at 0.7 ns, (e) recovering membrane integrity at 0.8 ns, (f) complete penetration at 1.7 ns.}
\end{figure}
Figure 4 shows a whole trajectory of the peptide and the membrane during a successful insertion, while Fig.5 shows the center-of-mass motion of the peptide.  Around 0.2 ns, the peptide begins to contact the membrane and starts insertion, until it finally comes to stay in the middle of the membrane. 
\begin{figure}
\includegraphics[width=.9\linewidth]{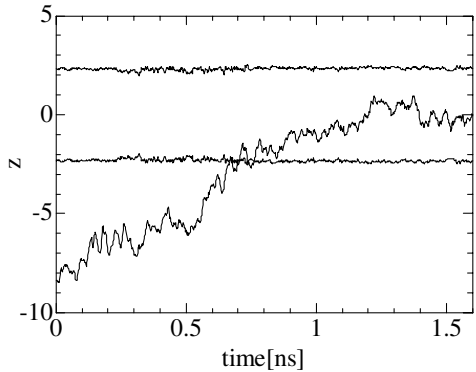}
\caption{ Time evolution of the reduced z-coordinate of the center-of-mass of the peptide.  Also shown are the positions of membrane surfaces (horizontal lines), which are represented by the centers of the hydrophilic groups of the lipids.}\end{figure}

  Typical peptide conformation after 1.0 ns of simulation is given in Fig.6a.  The hydrophilic loops are translocated to the other side across the membrane and the hydrophobic blocks are nearly stretched perpendicular to the membrane, which is a typical conformation of the membrane peptide.  The hydrophilic atoms of the peptide show nearly symmetric distribution (Fig. 6b), which indicates that about half of the hydrophilic blocks, two out of six in this case, are translocated across the membrane.  We should here remark that the number of blocks that successfully penetrate the membrane is not always half the total hydrophilic blocks. Sometime only a small fraction of the blocks translocate the membrane resulting in quite asymmetric distribution.  It is even the case when the peptide only adheres to the one side of the membrane and shows no sign of translocation.  
  
\begin{figure}
\includegraphics[width=.9\linewidth]{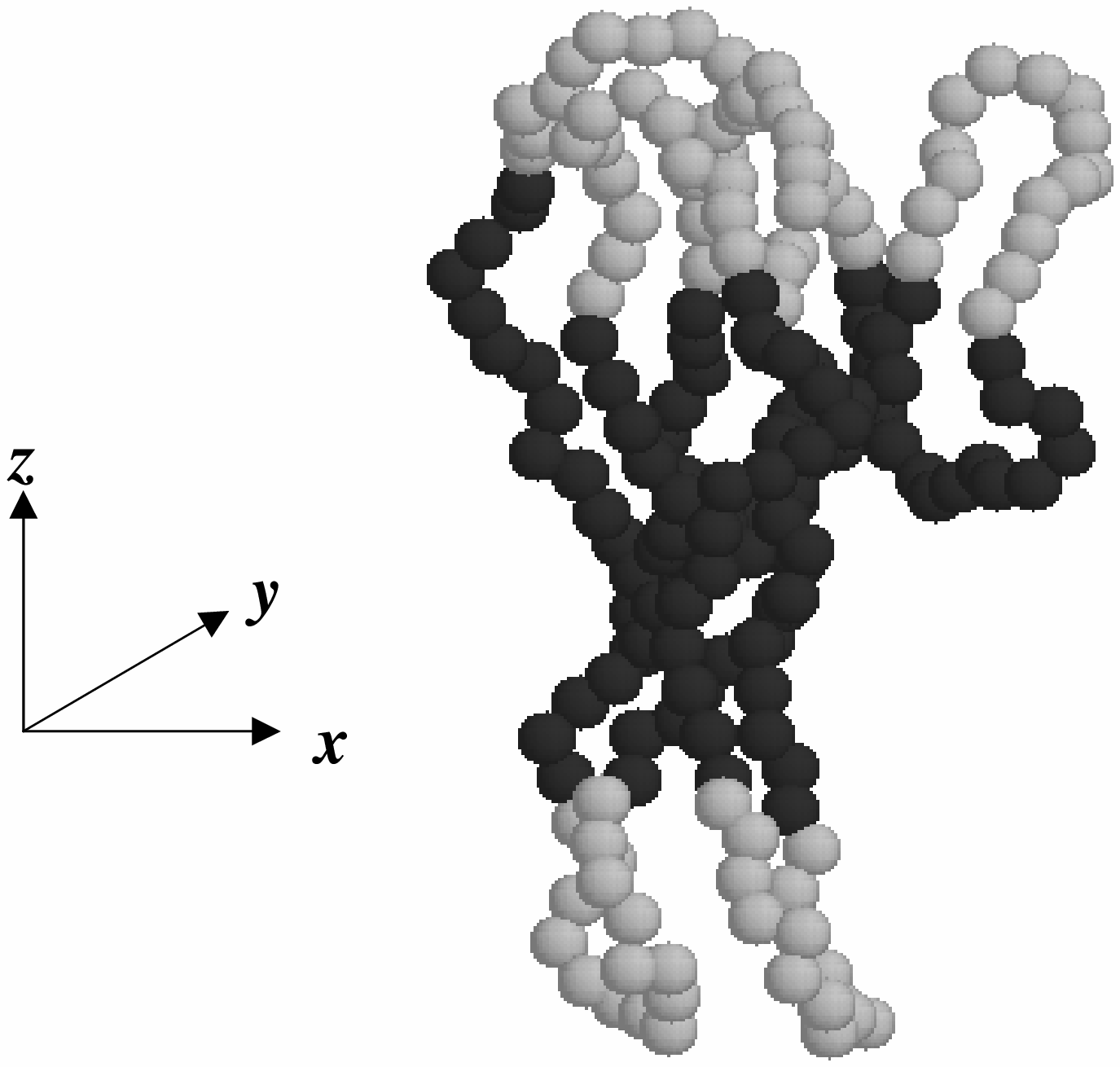}
\includegraphics[width=.9\linewidth]{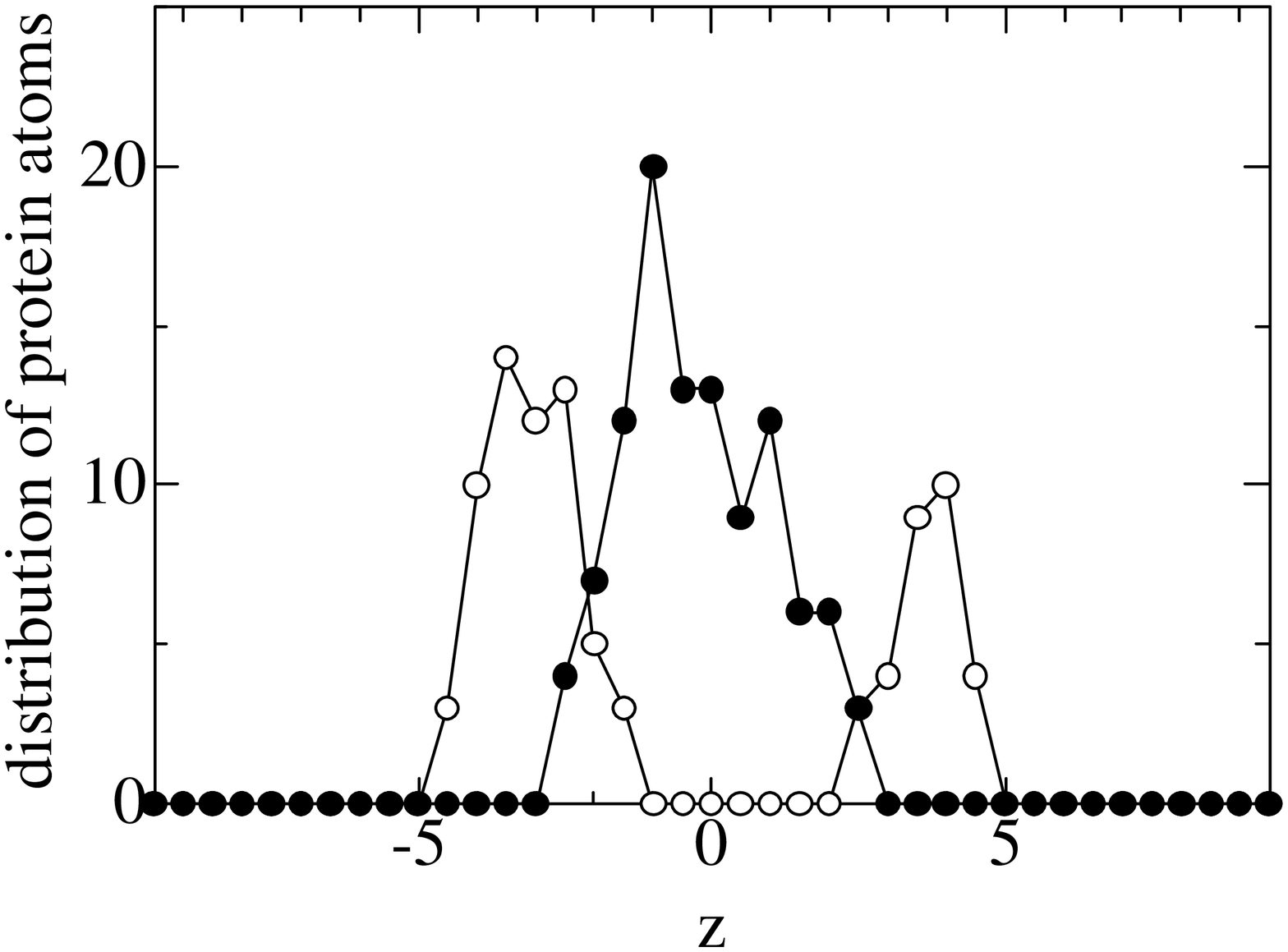}
\caption{ (a) Typical peptide conformation that has succeeded in the translocation, and (b) distributions of the component hydrophilic($\circ$) and hydrophobic($\bullet$) atoms along the z-axis. }
\end{figure}

 Lipid chains in the bilayers are known to show active diffusion within the layer, and with much less frequency between the layers; the later mode of chain diffusion is accompanied by the inversion of the chain head and ta
il and it is called flip-flop motion.  The frequency of the flip-flop motion is usually very low, but recent fluorescence experiment has revealed that it is greatly enhanced during the process of peptide insertion \cite{matsuzaki-1,matsuzaki-2}.  We here monitored the rate of chain inversion (flip-flop); we counted thefrequency of the chain inversion that the head-to-tail vectors of the chains change the sign of their z-component perpendicular to the membrane.  Fig.7 clearly shows that the flip-flop motion shows marked increases at the onset of peptide contact and insertion around 0.2 ns.  Most of the chain inversion is coupled withthe formation of the pore, the surface of which is covered with lipids pointing their hydrophilic heads perpendicularly toward the water phase within the pore. \begin{figure}
\includegraphics[width=.9\linewidth]{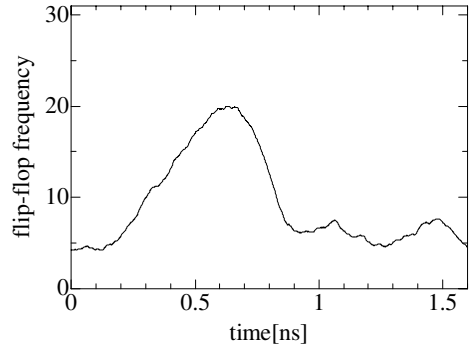}
\caption{  Total frequency of the flip-flop motions of the lipids within 2 ps interval; the data was averaged over $\pm50$ ps interval.  Marked increases are noticed during the contact and insertion of the peptide. }
\end{figure}
The driving force for the insertion and translocation of peptides has in general many origins, and most of them require metabolic energy for specific machinery \cite{cell}.  Even the short peptides which insert spontaneously can be driven by both energy and entropy.  We here monitored the energy during the insertion process and decomposed it into characteristic interaction energies: interactions between (b) lipid tails and protein H-beads@$E_{tail-H}$, (c) lipid heads and protein P-beads $E_{head-P}$, (d) lipid tails $E_{tail-tail}$, (e) protein H-beads $E_{H-H}$, (f) protein P-beads and water $E_{P-water}$, as well as (a) total interaction energy $E_{total}$ (Fig. 8).    Most conspicuous is the decrease in $E_{tail-H}$ by the insertion of the peptide's hydrophobic block into the hydrophobic membrane core.  This energy decrease is, however, counterbalanced both by the increase in $E_{tail-tail}$  (during the initial insertion up to 0.5 ns) which indicates the disordering in the membrane core region, and by the increase in $E_{H-H}$ (during the late stage of insertion after 0.5 ns) which is due to a slightly relaxed chain packing of the hydrophobic blocks of the peptide in the hydrophobic environment in the membrane core.  The overall energy change is only slightly negative, and this suggests the possible importance of the entropic contribution to the insertionprocess.  

\begin{figure*}
\includegraphics[width=.9\linewidth]{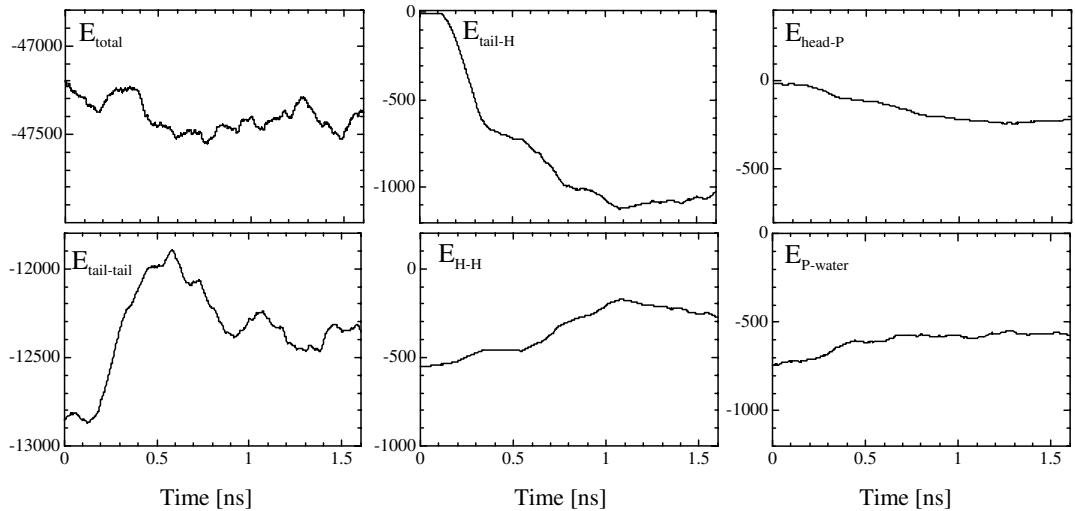}
\caption{Energy changes during the insertion process, and the decomposition into energies of different types of interactions; (a) total interaction energy , (b) the interaction between lipid tails and protein H-beads , (c) between lipid heads and protein P-beads  , (d) between lipid tails  , (e) between protein H-beads , and (f) between protein P-beads and water . The energy scale is the reduced unit, which is obtained by divided by . }  
\end{figure*}
 
Taking all these discussions into account, the mechanism of spontaneous insertion of short peptide-like polymers can be described as follows.  A free amphiphilic block-copolymer in water forms a unimolecular micelle with a collapsed hydrophilic core and extended hydrophilic arms, though real membrane peptides are know to be difficult to maintain such states in water due to their high hydrophobicity of the $\alpha$-helix forming stems.  When the polymer approaches the bilayers, flexible hydrophilic arms give way to the hydrophobic core in contacting the membrane.  Strong interactions between the core and the lipid tails perturb the membrane integrity resulting in a pore formation. The hydrophobic barrier to the peptide's translocation across the membrane is thus alleviated, and some of the hydrophilic blocks succeed in penetrating the membrane.  Such molecular picture may be a little bit far from conventional ones obtained from other molecular simulations.  But nice correspondence with recent experiments seems to support the present scenario.  Furthermore, it is quite reasonable to consider that the amphiphilic lipid molecules of the membrane act as surfactants for the hydrophobic core of the peptide forming larger micelle.  Thus the lipids and the peptide are required to perform highly cooperative actions during initial insertion, and through considerable reorganization of the lipids-peptide complex the peptide molecule completes the insertion into the membrane.

\begin{acknowledgments}
Present work was supported by the Grant-in-Aid of Scientific Research on Priority Areas, "Mechanism of Polymer Crystallization" (No. 12127206), from the Ministry of Education, Science, and Culture, Japan. 
\end{acknowledgments}

\end{document}